\documentclass{iaus}
\usepackage{graphicx}
\usepackage{graphics}
\title[Triggered collapse model confront observations] 
{Strongly triggered collapse model confront observations}

\author[Hennebelle et al.]   
{Patrick Hennebelle$^1$,
 Arnaud Belloche$^2$, Philippe Andr\'e$^3$  \and Anthony Whitworth$^4$}

\affiliation{$^1$ Laboratoire de Radioastronomie Millim\'etrique
, UMR 8112 du CNRS, \'Ecole Normale Sup\'erieure et Observatoire de Paris, 
24 rue Lhomond 75231 Paris Cedex 05 France 
\break patrick.hennebelle@ens.fr \\
$^2$ Max Planck-Institut fur Radioastronomie, Auf dem Hugel
69, 53121 Bonn, Germany \break email: belloche@mpifr-bonn.mpg.de \\[\affilskip]
$^3$ Service d'Astrophysique, CEA/DSM/DAPNIA, C.E. Saclay, 91191 Gif-sur-Yvette Cedex, France  email: pandre@cea.fr \\
$^4$ School of Physics \& Astronomy, Cardiff University, 5 The Parade, Cardiff CF24 3YB, Wales, UK  email: ant@astro.cf.ac.uk}

\pubyear{2004}
\volume{xxx}  
\pagerange{119--126}
\date{?? and in revised form ??}
\jname{Proceedings Title IAU Symposium}
\editors{A.C. Editor, B.D. Editor \& C.E. Editor, eds.}
\begin{document}

\maketitle

\begin{abstract}
Detailed modelling of individual protostellar condensations, 
is important to test the various theories. Here we present 
comparisons between strongly induced collapse models with 
one young class-0, IRAS4A in the Perseus cloud and one 
prestellar cloud observed in the Coalsack molecular cloud.
\end{abstract}

\firstsection 
\section{Introduction}

Triggered star formation has been proposed since many years (see 
e.g. ) as an important mode of star formation. Indeed many sources 
of star formation triggering have been proposed and are discussed
in this volume, namely turbulence, supernovae remnants, ionisation fronts,
stellar outflows or large scale collapse. 
All of these processes ought to play a r\^ole 
in inducing star formation and probably have  their own signatures. 
In order to be able to identify and to quantify their relative 
importance, it is necessary to study in great details protostellar 
condensations in which the collapse has been externally induced, 
looking for signatures of such violent triggering. 

Here we present detailed comparisons between induced collapse models and 
2 observed sources in which external triggering may have taken place, 
namely IRAS4A a well observed class-0 condensation located in the 
NGC1333 complex and the G2 globule located in the Coalsack molecular cloud.

In section 2, we first describe the models with which comparison 
will be performed. In section 3, comparison with IRAS4A is presented 
whereas comparison with the G2 globule is described in section 4. 

\section{Spontaneous versus induced collapse}

\begin{figure}
\includegraphics[width=8cm]{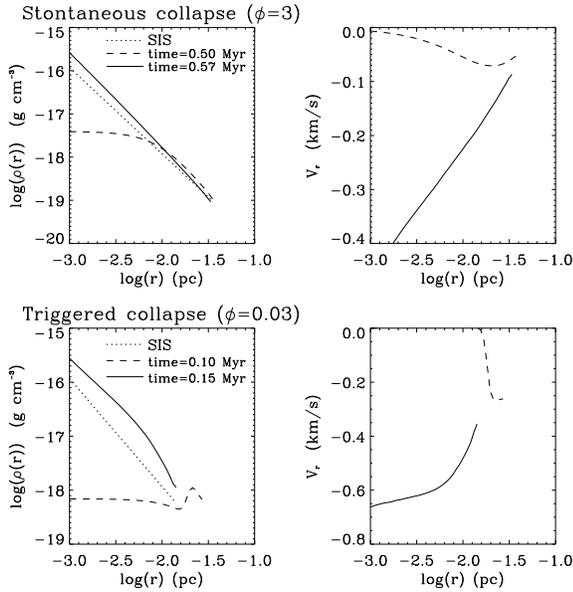}
  \caption{Density and velocity fields in the prestellar and class-0 
phase for a spontaneous  and a strongly triggered collapse.}
\label{fig:champ}
\end{figure}

In order to compare in details, observations of protostellar cores with 
theoretical models, it is necessary to construct a class of models 
which on one hand is realistic enough but on the other hand sufficiently 
simple to be described by few parameters that can be easily varied. 
It is for example unlikely  or at least computationally expensive 
to find in large scale turbulence simulations, a core
 which fits sufficiently well an observed object to allow very detailed 
comparisons. Our approach is as described 
in Hennebelle et al. (2003, 2004). We setup initially a stable 
Bonnor-Ebert sphere and we increase the external pressure at a given rate. 
In order to quantify the rapidity of the pressure increasement, we define
$\phi = (R_c/C_s) / (P/\dot{P})$ where $R_c$  is the cloud radius, 
$C_s$ the sound speed, $P$ the pressure and $\dot{P}$ its time derivative.
$\phi$ represents therefore the ratio of the sound crossing time over 
the typical  pressure increasement time. Large values of $\phi$ correspond
to slow pressure increasement and describe a spontaneous collapse whereas
for small values of $\phi$, the collapse is strongly externally triggered.
The calculations have been done with the SPH technique. 

Figure~\ref{fig:champ} shows the density and the velocity field 
in the equatorial plan before and 
after protostar formation for a spontaneous collapse ($\phi \simeq 3$)
and for a strongly triggered collapse ($\phi \simeq 0.03$). 
In the first case, the density appears to be close to a Bonnor-Ebert
sphere density in the prestellar phase, even when the cloud has become
unstable, and close to the density of the singular isothermal sphere
(SIS)  during the 
class-0 phase. The velocity remains subsonic during the prestellar phase 
and in the core outer part in the class-0 phase. It becomes more and more 
supersonic as the collapse proceeds, in the inner part. The situation is 
much different for $\phi=0.03$. The velocity is  supersonic 
in the outer part of the core during prestellar phase and everywhere during 
class-0 phase. The density is also very different from the previous case. 
It is higher in outer part than in the center. Indeed the strong pressure 
increasement has launched a compression wave that propagates indwards. 
During the class-0 phase, the density is significantly denser than the SIS
density. All of these features appear therefore to be characteristic 
of externally induced collapse.

\section{Comparison with the young class-0 IRAS4A}

\begin{figure}
\includegraphics[width=8cm]{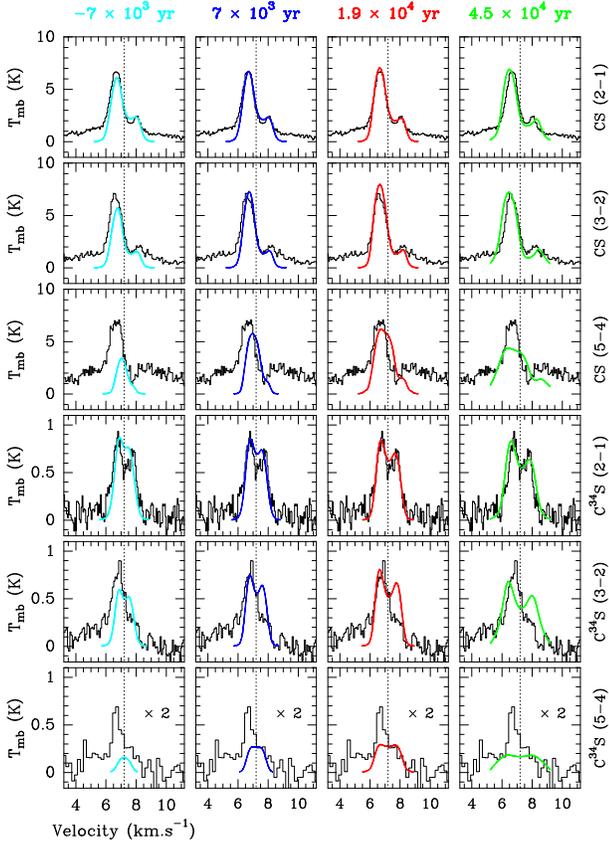}
  \caption{Comparaison between observed spectra and synthetic 
spectra obtained from the strongly induced collapse at various 
time.}
\label{fig:compar}
\end{figure}

NGC1333 is a well studied regions of the Perseus cloud. Several 
outflows have been detected (Knee \& Sandell 2000), at least one of them 
is pointing towards IRAS4A making the possibility of induced collapse 
for this source rather plausible. IRAS4A has been observed  
in various molecular lines by Di Francesco et al. (2001) and in the 
continuum by Motte \& Andr\'e (2001). Di Francesco et al. (2001) 
inferred supersonic velocity (0.5-1 km/s) whereas Motte \& Andr\'e (2001)
observed density up to 10 times the SIS density. As recalled 
in previous section, both are signatures of induced collapse. 
This has been confirmed by  observations done recently by 
Belloche et al. (2006). 

In order to confirm the scenario of induced collapse for IRAS4A 
 and to set accurate constraints, we have 
calculated synthetic spectra in the available lines, using 
the radiative transfer code described in Belloche et al. (2002) for 
the model $\phi=0.03$ presented in  the previous section.
Figure~\ref{fig:compar} shows the comparison between the data and 
the synthetic profiles at 4 time steps before and after protostellar
formation. The comparison reveals that good agreement 
can be obtained for almost all lines at time $1.9 \times 10^4$ yr
whereas at earlier times synthetic lines are generally too narrow and 
at later times, they are generally too broad.  As can be seen various 
features are not well reproduced even at time $1.9 \times 10^4$ yr. For example
the large wings seen in CS(2-1) are most probably due to the outflow 
launched by IRAS4A which is not taken into account in our modelling. 
It is also seen that C$^{34}S(5-4)$ is poorly reproduced. We 
speculate that the reason for this disagreement is that this line being 
optically thin, it traces the outer part of the core. The asymmetry of the 
lines may reveal that the collapse on large scales is indeed not symmetric.
 
We conclude that IRAS4A can be reasonably well reproduced by our model
although further refinements are highly wishable.

\section{Comparison with the prestellar G2 globule of Coalsack}

\begin{figure}
\includegraphics[width=12cm]{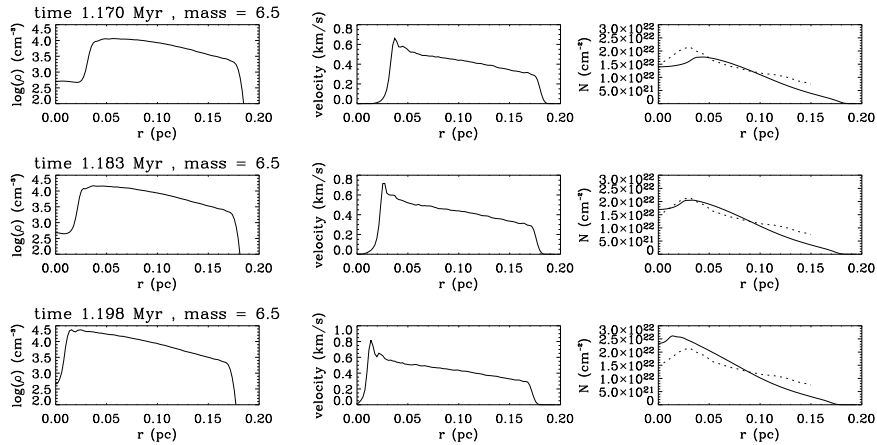}
  \caption{Density, velocity and column density of the 
 collapsing prestellar cloud (solid lines) and column density 
inferred from observations (dotted line) of Lada et al. (2004).}
\label{fig:ring}
\end{figure}

Lada et al. (2004) have recently observed in the Coalsack molecular cloud a 
core which presents an unusual ring-like structure around its centre. The 
column density appears therefore to decrease indwards when one appoaches
the core centre. Qualitatively this feature may have similarity with 
the compression wave shown in figure~\ref{fig:champ}. In our model 
the compression wave is a dense shell which in projection could be seen 
as a ring.  To test the validity of this scenario, we have  compared 
the column density observed by Lada et al. (2004) with the column density of 
our model at various time steps (Hennebelle et al. 2006). 
Figure~\ref{fig:ring} displays the density, velocity and column density 
fields at 3 time steps as well as the column density observed by Lada et al. 
(2004). The column density of the second case appears to be in good 
agreement with these observations. However no conclusion should be drawn 
until kinematical data are available for this core (only one spectrum has 
been observed). Indeed  we have calculated synthetic spectra, under 
optically thin assumption, towards various positions. Our model 
predicts that toward the centre, the lines should be strongly splitted.  
This prediction should be easy to test and we look forward to 
future observations of this peculiar object.



{}

\end{document}